Factors and moderators of ageism:
An analysis using data from 55 countries in the World Values Survey Wave 6

*Keisuke Kokubun[1]

[1]Graduate School of Management, Kyoto University, Kyoto, Japan
* kokubun.keisuke.6x@kyoto-u.jp

Abstract
Today, as the aging of the world accelerates, it is an urgent task to clarify factors that prevent ageism. In this study, using hierarchical multiple regression analysis of data from 40,869 people from 55 countries collected in the World Values Survey Wave 6, we showed that after controlling for demographic factors, stereotypes, a hungry spirit, and male chauvinism are related to ageism, and that altruism, trust within and outside the family, and trust in competition moderate the relationship between the independent and dependent variables. Furthermore, data from Japan, which has the highest aging rate and aging speed in the world, showed that these moderation relationships are moderated.

Introduction
Ageism is the process of systematic stereotyping and discrimination against people because of their age (Butler, 1969). Research on ageism has long been dominated by demographic studies, with few studies focusing on cultural aspects. In the latter case, studies are showing that stereotypes are related to ageism (Nelson, 2016; Swift et al., 2017). More recently, there have been studies showing that certain types of hungry spirit and masculinity, such as the pursuit of financial success and toughness, are related to ageism (Hövermann & Messner, 2023; Ng & Lim-Soh, 2021). However, little is known about the factors that moderate the relationship between these factors and ageism. Previous research suggests that in addition to altruism directed toward society at large, trusting relationships within and outside the family could be potential moderators (Hagestad & Uhlenberg, 2005; McPherson et al., 2001; Newman, Faux, & Larimer, 1997; Swift et al., 2017). In addition, if we apply the theory of the "marketized mentality" (Hövermann & Messner, 2023), it is thought that in a society where markets are developed and outlets exist, the power of the hungry spirit toward ageism is weakened.

Therefore, in this study, using the large-scale attitude data collected in the World Values Survey Wave 6 (Inglehart et al., 2014), and using hierarchical multiple regression analysis that controlled for demographic factors, we show that stereotypes, hungry spirit, and male chauvinism are related to ageism, and that altruism, trust within and outside the family, and trust in competition moderate the relationship between the independent and dependent variables. In addition, we show that the national character of Japan, where both the aging and aging rates are by far the highest, moderates the moderation relationship between these variables.

Review of previous research and presentation of hypotheses
Independent variables
In light of previous research, this study shows that stereotypes, hunger, and male chauvinism are related to ageism.

Stereotypes. Like other prejudices, ageism is based on many stereotypes (Nelson, 2016). Stereotypes about older people include those about physical and cognitive functioning, creativity, skills, productivity, burden on family and society, illness, frailty, dependency, asexuality, and loneliness (Swift et al., 2017). Therefore, stereotypes are thought to drive ageism. The following hypotheses are derived:

H1. Stereotypes are positively related to ageism.

Compared to stereotypes, it may be more difficult to intuitively imagine that the two things the author will mention next, namely, the hunger for success and male chauvinism, are linked to ageism. For this reason, empirical research on these issues has only recently begun to look into them, and the number is limited.

Hungry spirit. An analysis using large-scale data collected in the World Values Survey Wave 6 (Inglehart et al., 2014), the same as this study, shows that people who have a strong "marketized mentality" that emphasizes becoming rich and being successful in society are more likely to perceive the older people as a burden on society. This is because people who are obsessed with money and success are more sensitive to the decline in their share due to the increase in the social burden of supporting older people (Hövermann & Messner, 2023). Therefore, the following hypothesis is derived.

H2. Hungry spirit is positively correlated with ageism.

Male chauvinism. A study of 20 English-speaking countries showed that national ageism,

assessed using a corpus of 8 billion words, correlates with Hofstede's cultural scale of "masculinity" (Ng & Lim-Soh, 2021). Masculinity is related to ageism because in a society that values competition and values the strong and successful, older people, who are the polar opposite, are easily perceived as weak (Ng & Lim-Soh, 2021). Previous research, for example, has described how older men are alienated from younger men in their prime in social clubs in heavy industrial areas of England, where the economy is supported by men's manual labor (Pain et al., 2000). Thus, the following hypothesis is derived:

H3. Male chauvinism is positively correlated with ageism.

Moderators This study shows that altruism, trust within and outside the family, and trust in competition moderate the relationships between the independent and dependent variables.

Altruism. It is not surprising that the altruism of people who want to contribute to society in some way mitigates the negative effects on older people.

H4. Altruism moderates the positive relationship between the independent variables and ageism.

Trust. Family relationships are qualitatively different from other networks, such as acquaintances and friends, in that they provide a forum for connections that transcend age. Past empirical studies have argued that family relationships promote knowledge acquisition and favorable evaluations of other age groups (Newman et al., 1997) and help prevent age-based segregation and discrimination (McPherson et al., 2001). On the other hand, previous studies have shown that high age homogeneity in extrafamilial relationships, such as acquaintances and friends, hinders interactions across age groups and the perspective-taking necessary for understanding others (Hagestad & Uhlenberg, 2005). Relatedly, recent studies have shown that positive evaluations of older people occur in the context of family and partnerships, religion and spirituality, and work, while negative evaluations of older people occur in the context of interactions with friends and acquaintances, leisure activities, and social activities (Swift et al., 2017). Thus, the following hypotheses are derived:

H5. Trust within the family will moderate, weakening the positive relationship between the independent variables and ageism.

H6. Trust outside the family will moderate, strengthening the positive relationship between the independent variables and ageism.

If, as Hövermann & Messner (2023) argue, a hunger mentality leads to ageism when opportunities to gain profits in the market are denied, then conversely, the existence of a market in which profits can be easily obtained is likely to weaken the relationship between the independent variables and ageism. In other words, in an efficient market in which individual achievements are likely to lead to financial success, it should be difficult for some older people to unfairly obtain wealth, and therefore older people are unlikely to be scapegoated. Classical economics argues that healthy markets resolve various irrationalities in the world (Smith, 1853). Therefore, the following hypothesis is derived.

H7. Trust in competition moderates the positive relationship between the independent variables and ageism.

However, these hypotheses ignore regional differences. For example, in Japan, where the aging rate and aging speed are by far the highest, moderation is likely different from other countries. When the author calculated the aging rate (proportion of the population aged 65 and over) and aging speed (change in aging rate over the past 10 years) for the WVS 6 survey year using population data from the United Nations (2024) and the World Bank (2024), the averages for 59 countries were 8.546% and 0.992% points, respectively, while Japan's scores were 23.601% and 5.796% points, respectively, placing it far ahead of the second-place country in both categories (for reference, the second-place aging rate was Germany at 20.569%, and the second-place aging speed was South Korea at 3.912%). Therefore, the following hypothesis is derived.

H8. Japanese national character moderates the moderation relationship between variables.

Method
Sample and Data Collection
The data used were taken from Wave 6 of the World Values Survey (Inglehart et al., 2014). The original data was collected between 2010 and 2014 and included 89,565 respondents living in 60 countries. From this, data on respondents aged 60 years and older and missing data were removed, leaving data on 40,869 people (21,683 men and 19,186 women) aged 16-59 years from 55 countries used for the analysis. The data on those aged 60 years and older were removed primarily to minimize the influence of a certain "sense of ownership" of people who are close to becoming "70 years and older" on the results. Additionally, setting "60 years and older" as an exclusion criterion was also partly because some countries with a small older population use the definition of older people "60 years and older" instead of "70 years and older" in the questionnaire, as described below.

Measures

Dependent variable Ageism was measured using two independent items: "Older people are a burden on society" ("Burden") and "Older people get more than their fair share from the government" ("Unfair"). Both were measured on a 4-point Likert scale ranging from 1 ("Strongly disagree") to 4 ("Strongly agree"), with the original scores reversed.

Independent variable:

Stereotypes. Measured as the average of three items: "People over 70: are seen as friendly," "People over 70: are seen as competent," and "People over 70: viewed with respect." However, in countries where there are few "People over 70," these questions are asked to be answered about "People over 60" (World Values Survey Association, 2015). Each item was measured on a 5-point Likert scale ranging from 0 ("Very likely to be viewed that way") to 4 ("Not at all likely to be viewed that way") with the original scores reversed. In other words, the more negative stereotypes you perceive, the higher your score.

Hungry spirit. This was measured with a single item: "It is important for this person to be rich; to have a lot of money and expensive things." The original score was reversed and the item was measured on a 6-point Likert scale ranging from 1 ("Not at all like me") to 6 ("Very much like me").

Male chauvinism. It was measured by the average Z-score of four items: "On the whole, men make better political leaders than women do," "A university education is more important for a boy than for a girl," "On the whole, men make better business executives than women do," and "When jobs are scarce, men should have more right to a job than women." Z-scores were used to eliminate differences in scale between items. The first three items were reversed and measured on a four-point Likert scale from 1 ("Strongly disagree") to 4 ("Strongly agree"). The last item, "When jobs are scarce, men should have more right to a job than women," was reversed and measured on a three-point Likert scale from 1 ("Disagree") to 3 ("Agree").

Moderator:

Altruism. Measured as the average of four items: "It is important to this person to do something for the good of society," "It is important to this person to always behave properly;

to avoid doing anything people would say is wrong," "Looking after the environment is important to this person; to care for nature and save life resources," and "Tradition is important to this person; to follow the customs handed down by one's religion or family." Each item was measured on a 6-point Likert scale ranging from 1 ("Not at all like me") to 6 ("Very much like me") with the original scores reversed.

Trust within the family. This was measured with a single item, "How much you trust: Your family." The original score was reversed and measured on a 4-point Likert scale ranging from 1 ("Do not trust at all") to 4 ("Trust completely").

Trust relationships outside the family. This was measured using the average of five items: "How much you trust: Your neighborhood," "How much you trust: People you know personally," "How much you trust: People you meet for the first time," "How much you trust: People of another religion," and "How much you trust: People of another nationality." Each item was reversed and measured on a 4-point Likert scale ranging from 1 ("Do not trust at all") to 4 ("Trust completely").

Trust in competition. This was measured with a single item, "Competition good or harmful." The original scores were reversed and the measure was on a 10-point Likert scale ranging from 1 ("Competition is harmful. It brings out the worst in people") to 10 ("Competition is good. It stimulates people to work hard and develop new ideas").

Control variables
Happiness. It is not unnatural to think that individuals' values and happiness influence their attitudes toward the elderly. There is evidence that values and happiness are related to the economic level of countries and individuals (Inglehart & Welzel, 2005; Stevenson & Wolfers, 2008). Evaluations of the elderly have also changed from more positive to more negative in line with global industrialization and lifestyle changes that have occurred over the past 200 years (Ng & Lim-Soh, 2021).

Emancipative & Secular Values. Liberal societies emphasize self-expression, autonomy, choice, and a humanistic spirit (Inglehart and Welzel, 2005). Additionally, people who are accustomed to such societies have higher preferences for self-expression, freedom of lifestyle, gender equality, and individual autonomy, and are more tolerant of the environment, foreigners, and gays and lesbians. On the other hand, people who are not accustomed to such societies emphasize economic and physical security, value trust, are intolerant of diversity, and have ethnocentric attitudes. Similarly, people who are accustomed to secular societies place less importance on religion, authority, and the traditional family, and are more accepting of divorce, abortion, euthanasia, and suicide. On the other hand, people who are accustomed to secular societies value tradition and family and believe that children should be proud of their parents and should always love and respect them, regardless of how they behave (Inglehart and Welzel, 2005). Therefore, it is believed that people who hold more liberal and secular values, i.e., stronger emancipative values and secular values, tend to have a more negative attitude toward older people who place importance on traditional rules and norms. Both the secular values and emancipative values consist of five variables with various scales, which are integrated using a somewhat complicated formula and then converted to values between 0 and 1. In this study, the already calculated values included in Wave 6 are used as is. For details on the variable composition and calculation method, please refer to Inglehart and Welzel (2005).

Sense of happiness. Applying the discussion in the previous paragraph, it is not surprising that people who perceive themselves as healthy, wealthy, and happy would have negative attitudes toward older people, who are at the opposite end of the spectrum, and who are physically and mentally frail. Therefore, the sense of happiness was measured as the average Z-score of five items: "Satisfaction with your life," "Satisfaction with the financial situation of household," "How much freedom of choice and control over own life," "Feeling of happiness," and "State of health." Of these, "Satisfaction with your life" and "Satisfaction with the financial situation of household" were measured on a 10-point Likert scale ranging from 1 ("Completely dissatisfied") to 10 ("Completely satisfied"), and "How much freedom of choice and control over own life" was measured on a 10-point Likert scale ranging from 1 ("No choice at all") to 10 ("A great deal of choice"). On the other hand, "Feeling of happiness" and "State of health" were measured on a 4-point Likert scale ranging from 1 ("Not at all happy") to 4 ("Very happy") and 1 ("Poor") to 4 ("Very good"), respectively, with the original scores reversed.

We used 54 country dummies as variables (for example, the Japan dummy is 1 for Japan and 0 for the rest. One randomly selected country, South Africa, was used as the baseline). Previous studies have shown that variables that indicate the burden of supporting older people, such as the aging rate (Hövermann & Messner, 2023) and public pension rates (Chen et al., 2023), are related to ageism. In addition, the aging rate has also been shown to be related to ageism, although the positive and negative correlations are not consistent (Chen et al., 2023; Hövermann & Messner, 2023). On the other hand, economic indicators such as the level of modernization (Hövermann & Messner, 2023), GDP per capita, and Gini coefficient (Chen et al., 2023) are not directly related to ageism, but they may affect ageism little by little over a long period (Ng & Lim-Soh, 2021). Taking into account the impact of these variables, as well as other yet unexplored economic, social, and cultural variables, on ageism, this study uses country dummies in the analysis as a measure reflecting such differences.

Other variables measured included "Highest educational level attained" on a nine-point scale ranging from 1 ("No formal education") to 9 ("University-level education, with degree"). "Scales of income" was measured on a ten-point scale ranging from 1 ("Lowest group") to 10 ("Highest group"). "Social class" was measured on a five-point scale ranging from 1 ("Lower class") to 5 ("Upper class") by reversing the original scores. "Supervisor" is a variable indicating whether the respondent supervises subordinates at work; the original scores were transformed to measure it on a two-point scale of 1 ("Yes") or 0 ("No"). Other variables measured included gender (male: 1, female: 0), age, marital status (married or in a common-law marriage: 1, otherwise 0), and country.

Analytical method:
Hierarchical multiple regression analysis was conducted using two dependent variables of ageism, namely "Burden" and "Unfair." First, control variables were entered in Step 1, main variables in Step 2, and interaction terms in Step 3. All statistical analyses have been performed using IBM SPSS Statistics Version 26 (IBM Corp., Armonk, NY, USA).

Analysis and findings
Before the hierarchical multiple regression analysis, an exploratory factor analysis (EFA) was conducted to extract items for constructing variables for the regression analysis. The criterion for factor extraction was an eigenvalue of 1 or more, and factor loadings were calculated after varimax rotation using the principal factor method. Items with factor loadings of less than 0.4 and more than 0.4 for multiple factors were then excluded, and factor analysis was conducted again using the same criterion. This process was repeated until there were no items with factor loadings of less than 0.4 and more than 0.4 for multiple factors. Here, we followed the ideas of Stevens (1992). Finally, a factor structure consisting of "male chauvinism" ($\alpha = 0.767$), "trust outside the family" ($\alpha = 0.752$), "happiness" ($\alpha = 0.687$), "altruism" ($\alpha = 0.717$), and "stereotyping" ($\alpha = 0.679$) was established. In this process, the item "How much you trust: Your family" ("trust within the family"), which does not load on a specific factor, was made into an independent variable consisting of one item. On the other hand, "Older people are a burden on society" ("Burden") and "Older people get more than their fair share from the government" ("Unfair") were loaded onto one common factor with several other variables, but they were treated as independent variables consisting of one item each because the reliability coefficient was below 0.6 (Taber, 2018; Van Griethuijsen et al., 2015), which is the standard used by several researchers. For the same reason, the following items were also treated as single variables: "It is important to this person to be rich; to have a lot of money and expensive things" ("Hungry spirit") and "Competition good or harmful" ("Trust in competition"). The following five items were used in the factor analysis but ultimately could not be used to form a factor and were therefore abandoned. "Companies that employ young people perform better than those that employ people of different ages", "Old people have too much political influence", "It is important for this person to think up new ideas and be creative; to do things one's own way", "Adventure and taking risks are important to this person; to have an exciting life", "Hard work brings success". Table 1 presents descriptive statistics.

Table 2 shows the results of the hierarchical multiple regression analysis. In Step 1, only the control variables were included. As a result, seven control variables, excluding marriage and social class, showed a significant correlation at the 5% level with either or both of the two dependent variables. Due to space limitations, the 54 country dummies are omitted from the table, but all dummy variables for "Burden" and most dummy variables for "Unfair" showed a significant correlation at the 0.1% level (available upon request). In Step 2, the main variables were added. For both "Burden" and "Unfair," the independent variables "Stereotype" ($\beta = 0.104$, p < 0.01; $\beta = 0.035$, p < 0.01), "Hungry spirit" ($\beta = 0.055$, p < 0.01; $\beta = 0.051$, p < 0.01), and "Male chauvinism" ($\beta = 0.185$, p < 0.01; $\beta = 0.144$, p < 0.01) showed a significant positive correlation at the 0.1% level. These support H1 to H3. All four moderators were significantly correlated with the dependent variables at the 0.1% level, except for "Trust outside the family" for "Unfair."

In Step 3, interaction terms were entered. First, in the model with "Burden" as the dependent variable, significant correlations were shown for Stereotype × Trust outside the family ($\beta = 0.015$, p < 0.1), Hungry spirit × Trust inside the family ($\beta = -0.018$, p < 0.01), Hungry Spirit × Trust outside the family ($\beta = 0.034$, p < 0.01), Hungry spirit × Confidence in competition ($\beta = -0.021$, p < 0.01), Male chauvinism × Altruism ($\beta = -0.013$, p < 0.1), Male chauvinism × Trust outside the family ($\beta = 0.021$, p < 0.01), and Male chauvinism × Confidence in competition ($\beta = -0.024$, p < 0.01). On the other hand, in the model with "Unfair" as the dependent variable, significant correlations were observed for Stereotype × Altruism ($\beta = -0.015$, p < 0.1), Hungry spirit × Trust inside the family ($\beta = -0.014$, p < 0.1), Hungry spirit × Trust outside the family ($\beta =$, p < 0.01), Hungry spirit × Confidence in competition ($\beta = 0.024$, p < 0.01), and Male chauvinism × Confidence in competition ($\beta = -0.016$, p < 0.1). These support H4 to H7.

TABEL 1. Descriptive statistics and correlation coefficients

| | | Mean | SD | 1 | 2 | 3 | 4 | 5 | 6 | 7 | 8 | 9 | 10 | 11 | 12 | 13 | 14 | 15 | 16 | 17 | 18 |
|---|---|---|---|---|---|---|---|---|---|---|---|---|---|---|---|---|---|---|---|---|---|
| 1 | Sex | 0.531 | 0.499 | | | | | | | | | | | | | | | | | | |
| 2 | Age | 36.700 | 11.557 | -0.012* | | | | | | | | | | | | | | | | | |
| 3 | Marriage | 0.631 | 0.483 | -0.007 | 0.350*** | | | | | | | | | | | | | | | | |
| 4 | Education | 6.020 | 2.210 | -0.015** | -0.108*** | -0.069*** | | | | | | | | | | | | | | | |
| 5 | Income | 5.010 | 2.058 | 0.010* | -0.054*** | 0.023*** | 0.249*** | | | | | | | | | | | | | | |
| 6 | Position | 0.321 | 0.467 | 0.123*** | 0.121*** | 0.100*** | 0.138*** | 0.148*** | | | | | | | | | | | | | |
| 7 | Social class | 2.715 | 1.006 | -0.001 | 0.004 | 0.042*** | 0.315*** | 0.452*** | 0.182*** | | | | | | | | | | | | |
| 8 | Secular values | 0.360 | 0.173 | 0.000 | -0.052*** | -0.110*** | 0.080*** | -0.034*** | -0.047*** | -0.055*** | | | | | | | | | | | |
| 9 | Emancipative values | 0.409 | 0.170 | -0.135*** | 0.017*** | -0.075*** | 0.207*** | 0.036*** | 0.040*** | 0.071*** | 0.346*** | | | | | | | | | | |
| 10 | Wellbeing | 0.000 | 0.666 | 0.011* | -0.109*** | 0.047*** | 0.100*** | 0.339*** | 0.101*** | 0.247*** | -0.183*** | 0.064*** | | | | | | | | | |
| 11 | Stereotype | 1.215 | 0.859 | -0.023*** | -0.024*** | -0.039*** | 0.005 | -0.052*** | -0.004 | -0.019*** | 0.188*** | 0.072*** | -0.132*** | | | | | | | | |
| 12 | Hungry spirit | 3.640 | 1.531 | -0.058*** | 0.122*** | 0.045*** | 0.061*** | -0.105*** | 0.010* | -0.017** | 0.033*** | 0.194*** | 0.026*** | 0.010 | | | | | | | |
| 13 | Male chauvinism | 0.000 | 0.765 | 0.241*** | -0.023*** | 0.047*** | -0.155*** | 0.009 | -0.039*** | -0.044*** | -0.140*** | -0.617*** | -0.076*** | -0.044*** | -0.234*** | | | | | | |
| 14 | Altruism | 4.515 | 0.961 | -0.002 | 0.021*** | 0.057*** | -0.038*** | -0.001 | 0.025*** | 0.026*** | -0.411*** | -0.221*** | 0.097*** | -0.159*** | -0.091*** | 0.132*** | | | | | |
| 15 | Trust inside the family | 3.778 | 0.527 | 0.032*** | 0.017*** | 0.075*** | 0.069*** | 0.078*** | 0.025*** | 0.085*** | -0.152*** | -0.058*** | 0.117*** | -0.095*** | 0.017*** | 0.016** | 0.116*** | | | | |
| 16 | Trust outside the family | 2.424 | 0.585 | 0.020*** | 0.066*** | 0.013** | 0.104*** | 0.109*** | 0.045*** | 0.080*** | -0.063*** | 0.096*** | 0.079*** | -0.024*** | -0.034*** | -0.011* | 0.013* | 0.198*** | | | |
| 17 | Confidence in competition | 3.900 | 2.637 | -0.047*** | -0.011* | -0.039*** | -0.048*** | 0.005 | -0.060*** | -0.056*** | 0.180*** | 0.058*** | -0.042*** | 0.048*** | -0.029*** | 0.005 | -0.140*** | -0.112*** | -0.018*** | | |
| 18 | Burden | 3.150 | 0.804 | -0.015** | 0.018*** | 0.024*** | 0.064*** | -0.020*** | 0.031*** | 0.050*** | -0.128*** | 0.013* | 0.027*** | -0.120*** | 0.119*** | -0.134*** | 0.077*** | 0.086*** | -0.026*** | -0.095*** | |
| 19 | Unfair | 2.860 | 0.878 | -0.015** | 0.040*** | 0.018*** | 0.086*** | -0.036*** | 0.000 | 0.031*** | -0.009 | 0.089*** | -0.039*** | -0.049*** | 0.140*** | -0.179*** | 0.022*** | 0.068*** | -0.014** | -0.074*** | 0.318*** |

Note: n = 40.869. *** p < 0.001, ** p < 0.01, * p < 0.05.

TABEL 2. Multiple regression analysis with ageism ("burden" and "unfair") as the dependent variable

|  | Step 1 Burden | Step 1 Unfair | Step 2 Burden | Step 2 Unfair | Step 3 Burden | Step 3 Unfair | Step 4 Burden | Step 4 Unfair |
|---|---|---|---|---|---|---|---|---|
| *Control* | | | | | | | | |
| Sex | 0.010* | 0.008 | -0.018*** | -0.015** | -0.016** | -0.014** | -0.016** | -0.014** |
| Age | -0.002 | -0.024*** | 0.004 | -0.019*** | 0.005 | -0.018*** | 0.005 | -0.018*** |
| Marriage | 0.007 | 0.000 | 0.008 | 0.001 | 0.008 | 0.000 | 0.008 | 0.001 |
| Education | -0.050*** | -0.044*** | -0.037*** | -0.033*** | -0.034*** | -0.032*** | -0.034*** | -0.032*** |
| Income | 0.032*** | 0.036*** | 0.023*** | 0.027*** | 0.021*** | 0.027*** | 0.021*** | 0.027*** |
| Position | -0.018*** | -0.020*** | -0.015** | -0.017*** | -0.016** | -0.017*** | -0.016** | -0.017*** |
| Social class | 0.009 | 0.002 | 0.005 | -0.001 | 0.004 | -0.001 | 0.004 | -0.001 |
| Secular values | 0.126*** | 0.056*** | 0.078*** | 0.022*** | 0.076*** | 0.021*** | 0.076*** | 0.021*** |
| Emancipative values | -0.050*** | -0.062*** | 0.047*** | 0.015* | 0.047*** | 0.014* | 0.047*** | 0.014* |
| Wellbeing | -0.012* | 0.020*** | 0.005 | 0.030*** | 0.004 | 0.029*** | 0.004 | 0.030*** |
| *Main variable* | | | | | | | | |
| Stereotype | | | 0.104*** | 0.035*** | 0.102*** | 0.034*** | 0.103*** | 0.034*** |
| Hungry spirit | | | 0.055*** | 0.051*** | 0.059*** | 0.051*** | 0.059*** | 0.051*** |
| Male chauvinism | | | 0.185*** | 0.144*** | 0.185*** | 0.142*** | 0.185*** | 0.142*** |
| *Moderator* | | | | | | | | |
| Altruism | | | -0.024*** | -0.022*** | -0.027*** | -0.023*** | -0.027*** | -0.023*** |
| Trust inside the family | | | -0.053*** | -0.034*** | -0.050*** | -0.032*** | -0.050*** | -0.033*** |
| Trust outside the family | | | 0.027*** | 0.007 | 0.024*** | 0.004 | 0.024*** | 0.005 |
| Confidence in competition | | | -0.035*** | -0.036*** | -0.033*** | -0.035*** | -0.033*** | -0.035*** |
| *Interaction (Stereotype)* | | | | | | | | |
| Stereotype×Altruism | | | | | 0.003 | -0.015** | 0.003 | -0.014** |
| Stereotype×Trust intside the family | | | | | -0.003 | 0.000 | -0.004 | 0.000 |
| Stereotype×Trust outside the family | | | | | 0.015** | -0.001 | 0.016** | 0.000 |
| Stereotype×Confidence in competition | | | | | 0.005 | -0.006 | 0.004 | -0.006 |
| *Interaction (Hungry spirit)* | | | | | | | | |
| Hungry spirit×Altruism | | | | | -0.010 | 0.001 | -0.010* | 0.001 |
| Hungry spirit×Trust inside the family | | | | | -0.018*** | -0.014** | -0.018*** | -0.014** |
| Hungry spirit×Trust outside the family | | | | | 0.034*** | 0.024*** | 0.034*** | 0.023*** |
| Hungry spirit×Confidence in competition | | | | | -0.021*** | -0.016** | -0.021*** | -0.016** |
| *Interaction (Male chauvinism)* | | | | | | | | |
| Male chauvinism×Altruism | | | | | -0.013** | 0.004 | -0.013** | 0.004 |
| Male chauvinism×Trust inside the family | | | | | 0.004 | -0.003 | 0.003 | -0.003 |
| Male chauvinism×Trust outside the family | | | | | 0.021*** | 0.010 | 0.021*** | 0.01 |
| Male chauvinism×Confidence in competition | | | | | -0.024*** | -0.020*** | -0.024*** | -0.020*** |
| *Interaction (Japan)* | | | | | | | | |
| Japan | | | | | | | -0.057*** | -0.013 |
| Stereotype×Japan | | | | | | | -0.006 | 0.004 |
| Hungry spirit×Japan | | | | | | | 0.007 | 0.002 |
| Male chauvinism×Japan | | | | | | | -0.018 | -0.004 |
| Altruism×Japan | | | | | | | 0.008 | -0.003 |
| Trust inside the family×Japan | | | | | | | 0.000 | 0.01 |
| Trust outside the family×Japan | | | | | | | -0.003 | -0.001 |
| Confidence in competition×Japan | | | | | | | -0.008 | -0.008 |
| Stereotype×Altruism×Japan | | | | | | | -0.002 | -0.002 |
| Stereotype×Trust intside the family×Japan | | | | | | | 0.005 | -0.002 |
| Stereotype×Trust outside the family×Japan | | | | | | | -0.008 | -0.013* |
| Stereotype×Confidence in competition×Japan | | | | | | | 0.013* | -0.001 |
| Hungry spirit×Altruism×Japan | | | | | | | 0.014 | 0.001 |
| Hungry spirit×Trust inside the family×Japan | | | | | | | -0.003 | 0.006 |
| Hungry spirit×Trust outside the family×Japan | | | | | | | 0.003 | 0 |
| Hungry spirit×Confidence in competition×Japan | | | | | | | -0.003 | -0.008 |
| Male chauvinism×Altruism×Japan | | | | | | | -0.017 | -0.004 |
| Male chauvinism×Trust inside the family×Japan | | | | | | | 0.007 | -0.004 |
| Male chauvinism×Trust outside the family×Japan | | | | | | | -0.001 | 0.001 |
| Male chauvinism×Confidence in competition×Japan | | | | | | | -0.005 | 0.004 |
| F | 78.109*** | 148.743*** | 98.102*** | 149.987*** | 86.441*** | 129.775*** | 70.511*** | 105.735*** |
| adj. $R^2$ | 0.108 | 0.188 | 0.144 | 0.206 | 0.148 | 0.207 | 0.148 | 0.207 |

Note: n = 40.869. *** $p < 0.001$, ** $p < 0.01$, * $p < 0.05$.

The analysis controlled for the following 55 country dummy variables.
Algeria, Azerbaijan, Argentina, Australia, Armenia, Brazil, Belarus, Chile, China, Colombia, Cyprus, Ecuador, Egypt, Estonia, Georgia, Germany, Ghana, Haiti, Hong Kong, Iraq, Japan, Jordan, Kazakhstan, Kuwait, Kyrgyzstan, Lebanon, Libya, Malaysia, Mexico, Netherlands, Nigeria, Pakistan, Palestine, Peru, Philippines, Poland, Qatar, Romania, Russia, Rwanda, Singapore, Slovenia, South Africa, South Korea, Sweden, Taiwan, Thailand, Trinidadand Tobago, Tunisia, Turkey, Ukraine, United States, Uruguay, Uzbekistan, Zimbabwe.

In Step 4, Japan and the independent variables, as well as the interaction terms between Japan and the moderators, were entered. First, in the model with "Burden" as the dependent variable, significant correlations were shown for Japan ($\beta = -0.057$, $p < 0.01$) and Stereotype × Confidence in competition × Japan ($\beta = 0.013$, $p < 0.5$). On the other hand, in the model with "Unfair" as the dependent variable, significant correlations were shown for Stereotype × Trust outside the family × Japan ($\beta = -0.013$, $p < 0.5$). These support H8.

To further understand the significance of the interaction terms, in Figures 1 to 12, the data were divided into two groups with high and low independent variables. The criterion for high and low is whether the score is 1 SD higher or lower than the average, following the recommendation of Aiken et al. (1991).

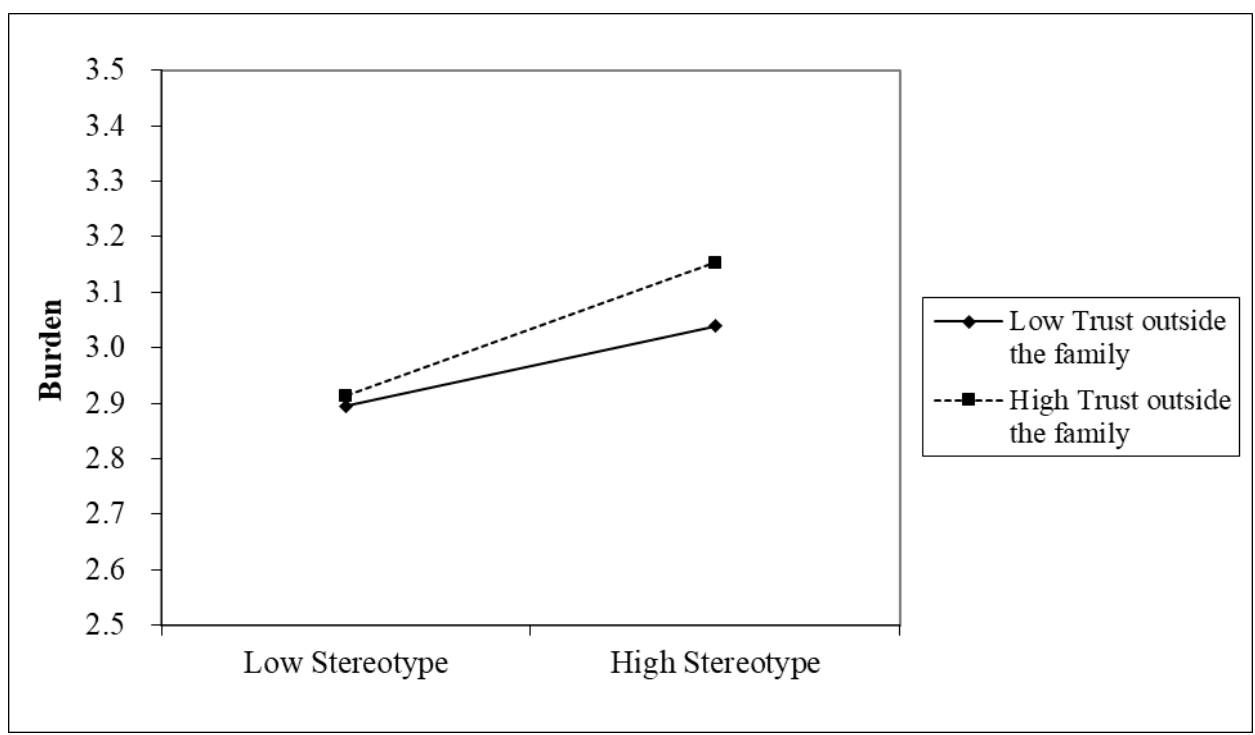

FIGURE 1 The moderating effect of trust outside the family between stereotype and burden

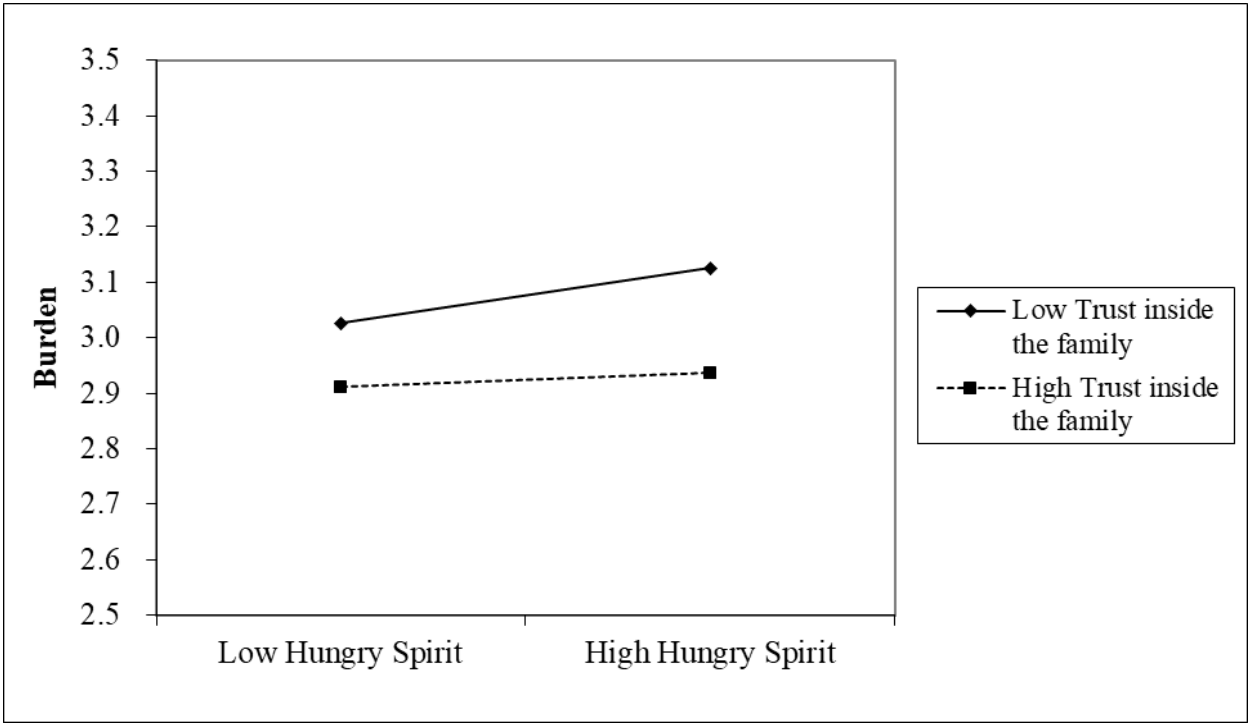

FIGURE 2   The moderating effect of trust inside the family between hungry spirit and burden

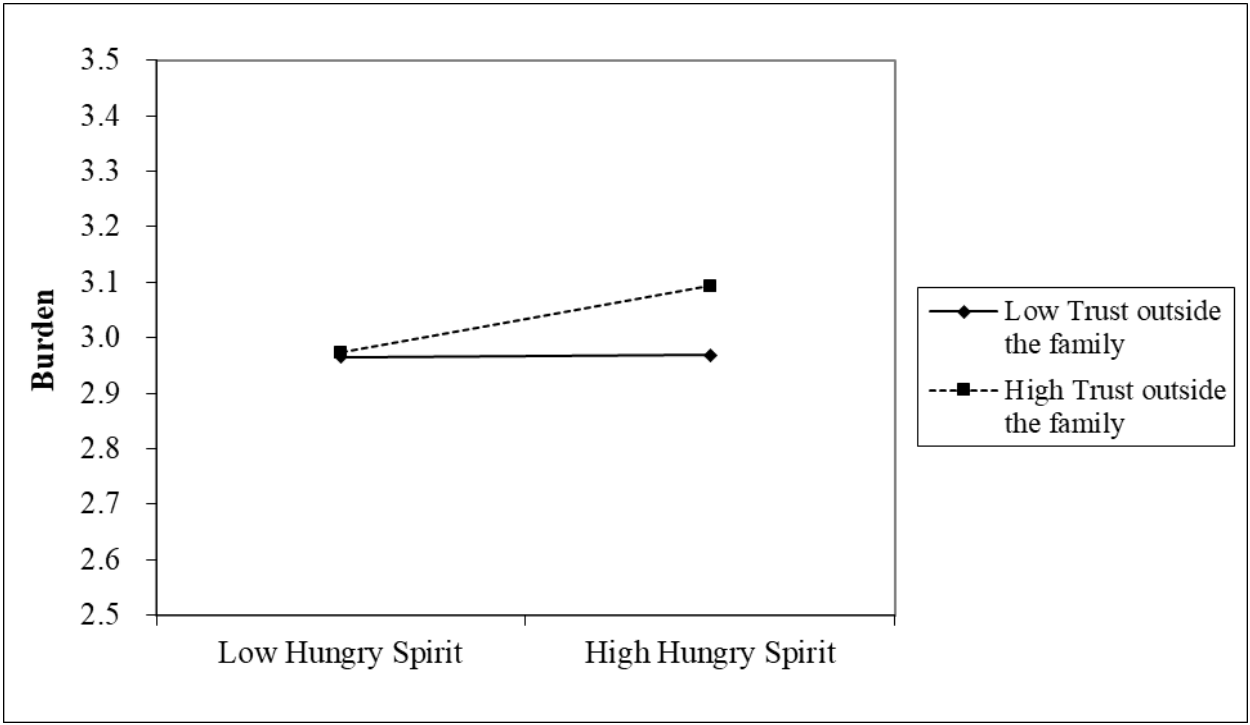

FIGURE 3   The moderating effect of trust outside the family between hungry spirit and burden

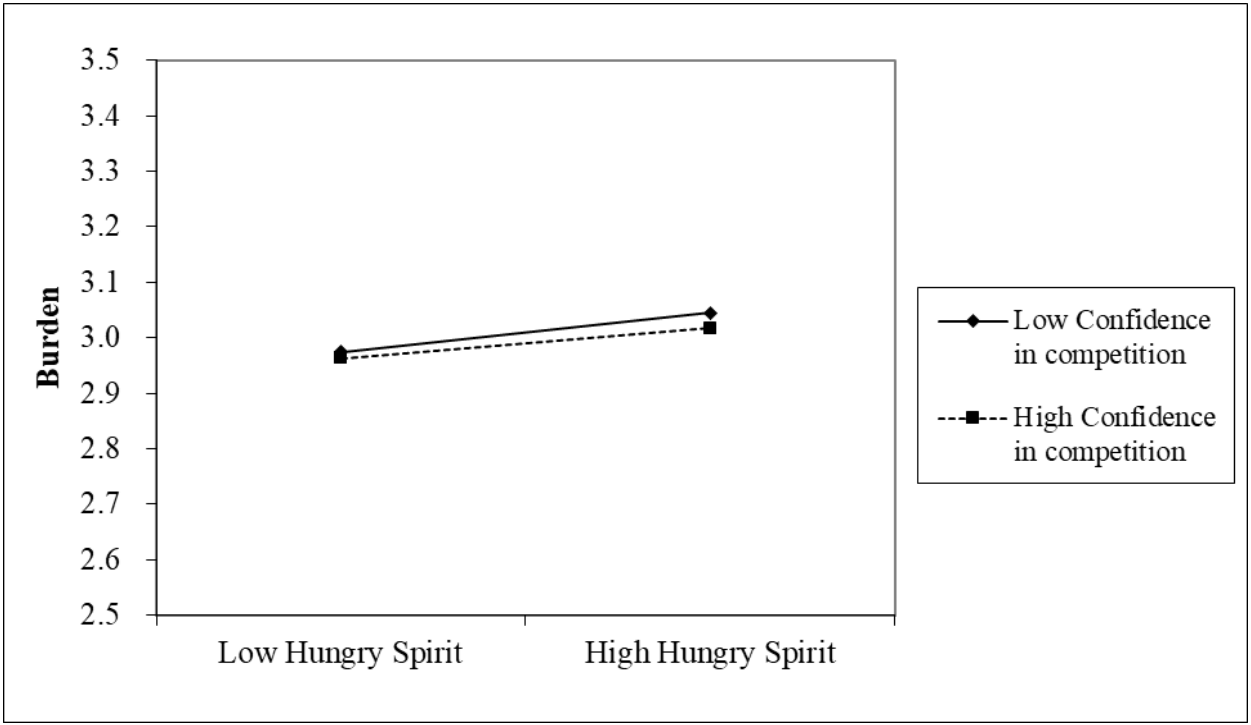

FIGURE 4   The moderating effect of confidence in competition between hungry spirit and burden

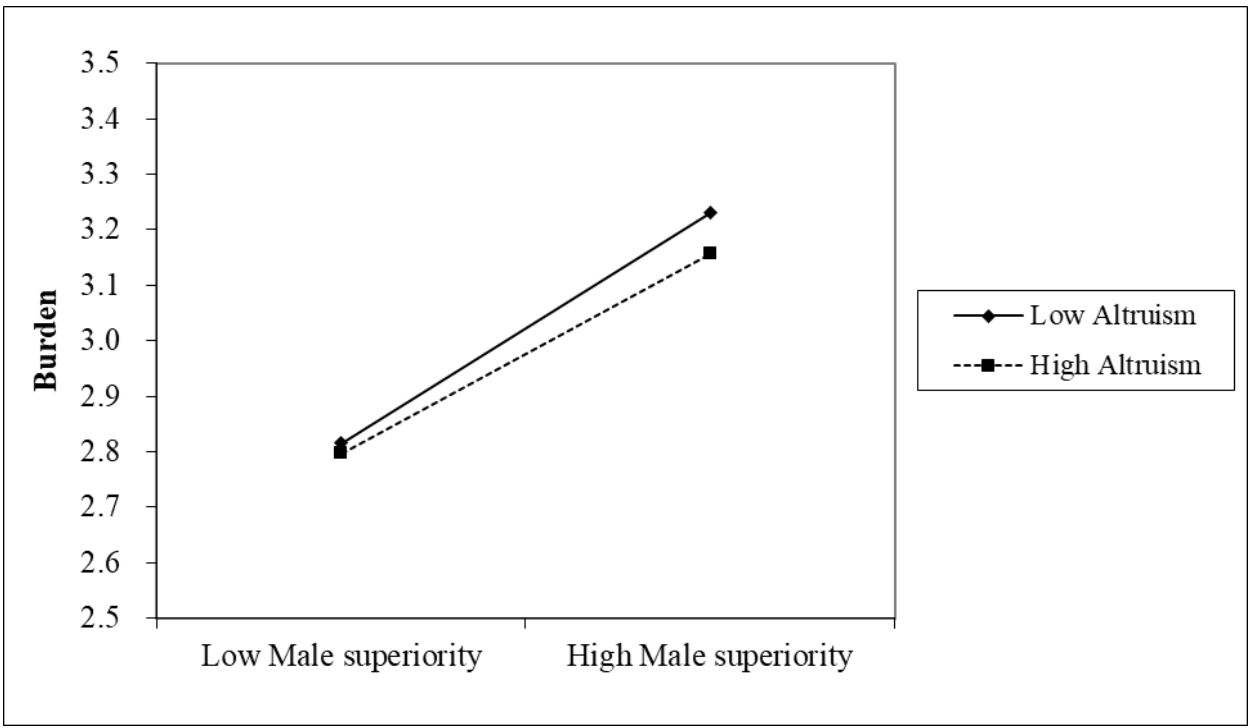

FIGURE 5   The moderating effect of altruism between male chauvinism and burden

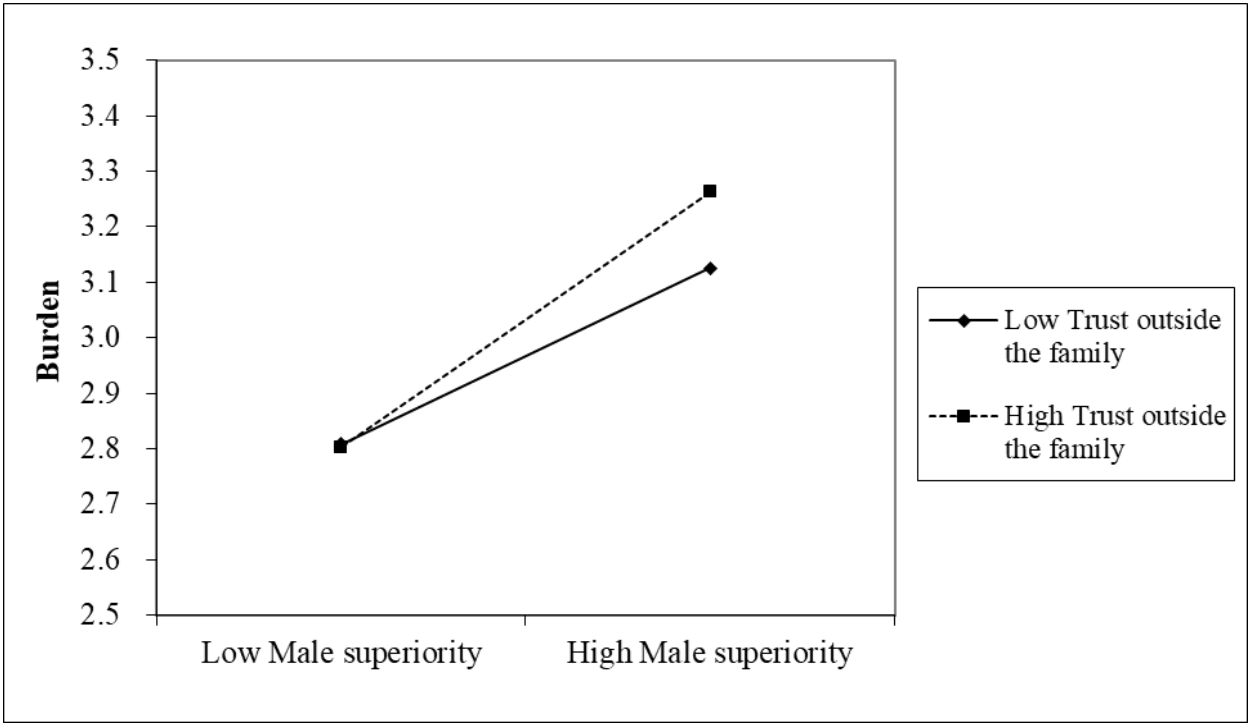

FIGURE 6　The moderating effect of trust outside the family between male chauvinism and burden

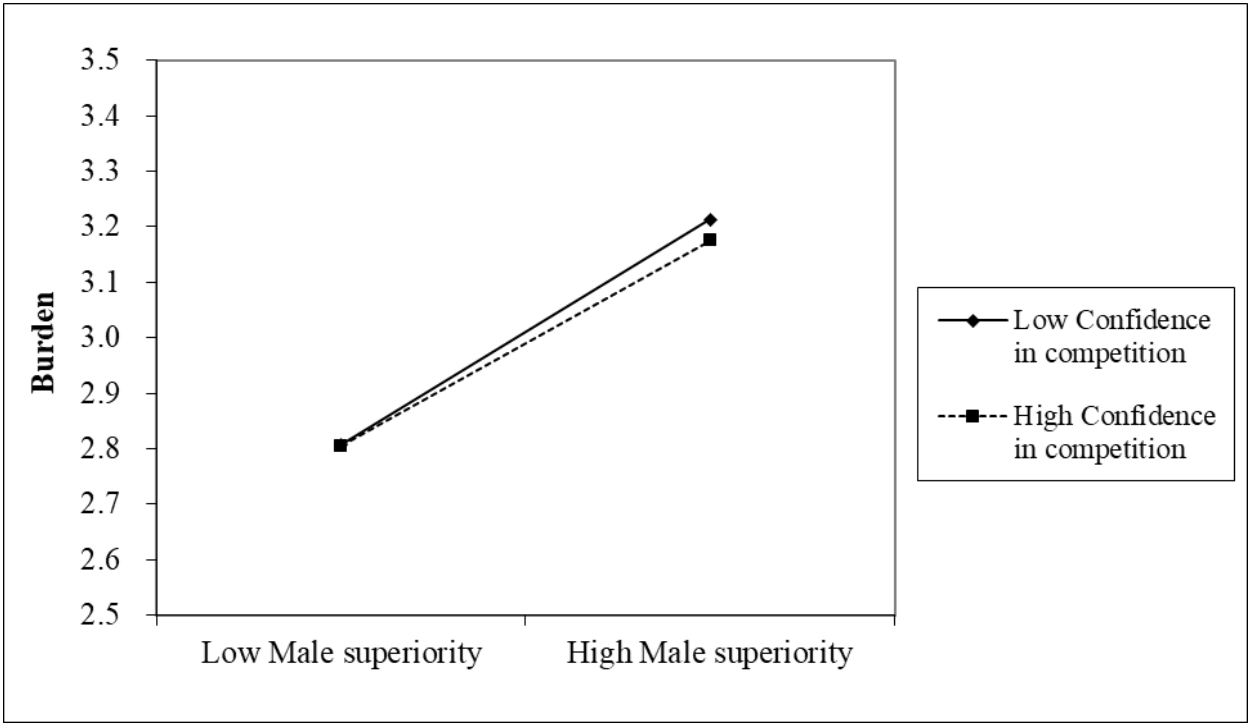

FIGURE 7　The moderating effect of confidence in competition between male chauvinism and burden

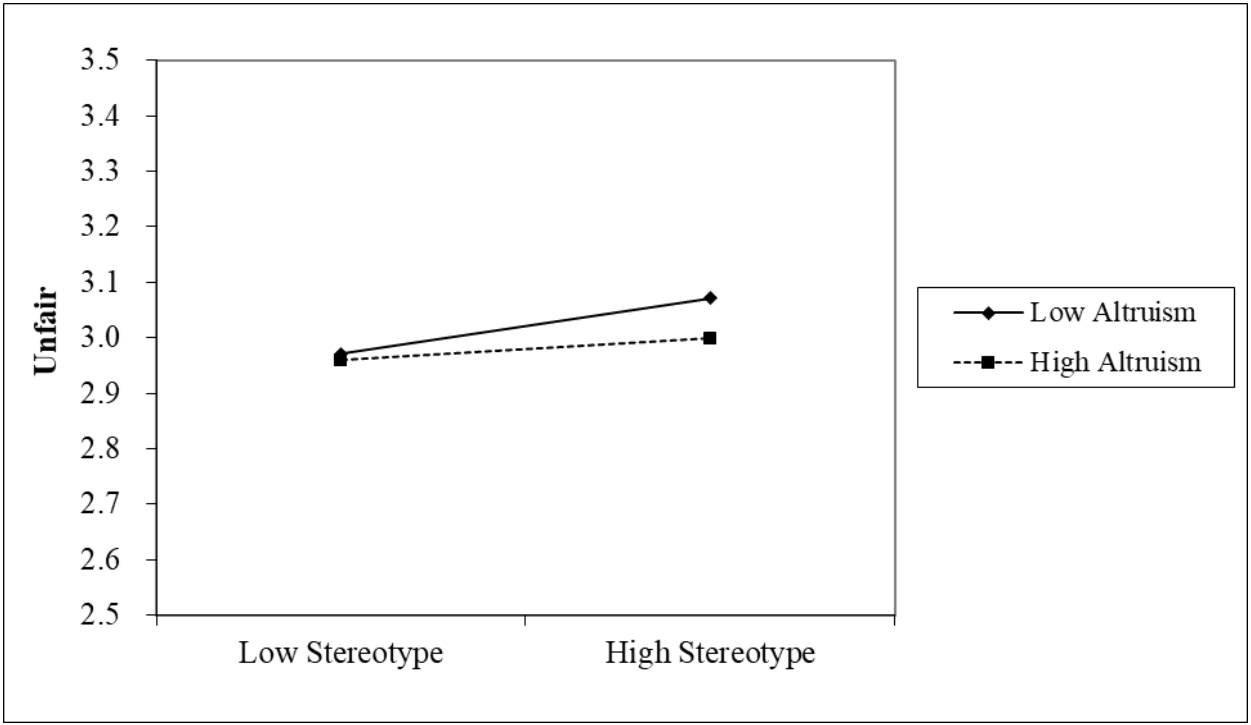

FIGURE 8  The moderating effect of altruism between stereotype and unfair

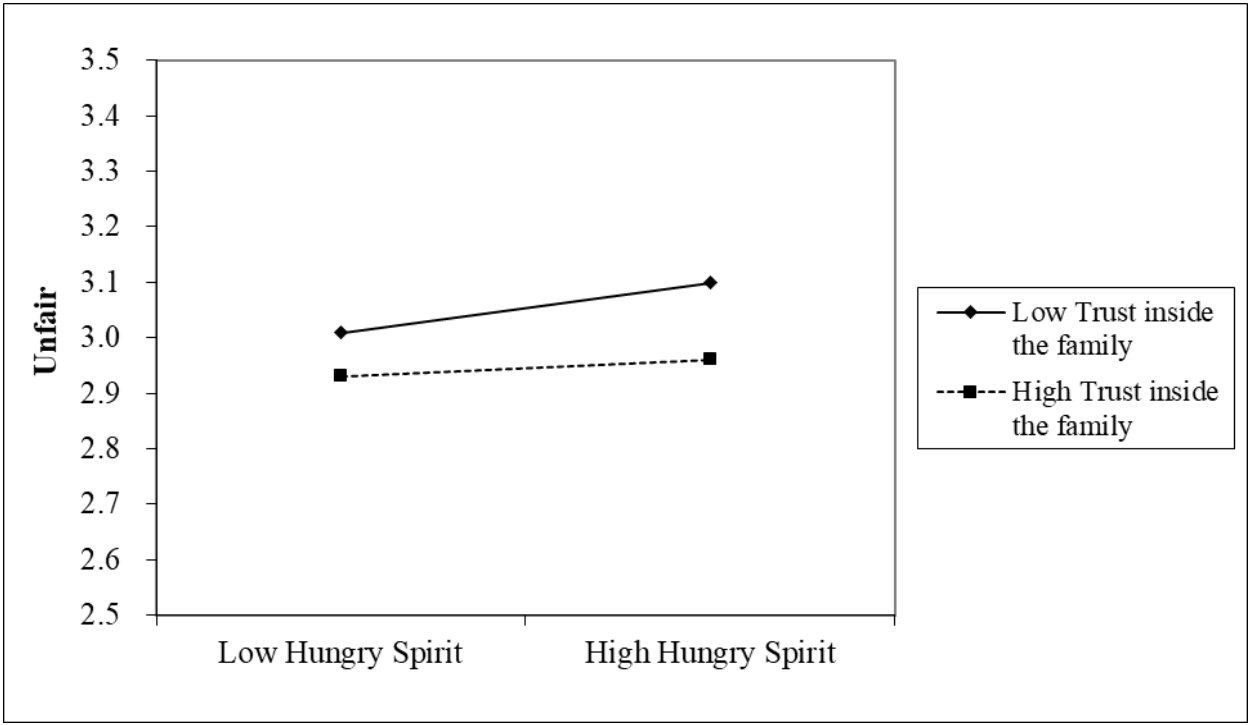

FIGURE 9  The moderating effect of trust inside the family between hungry spirit and unfair

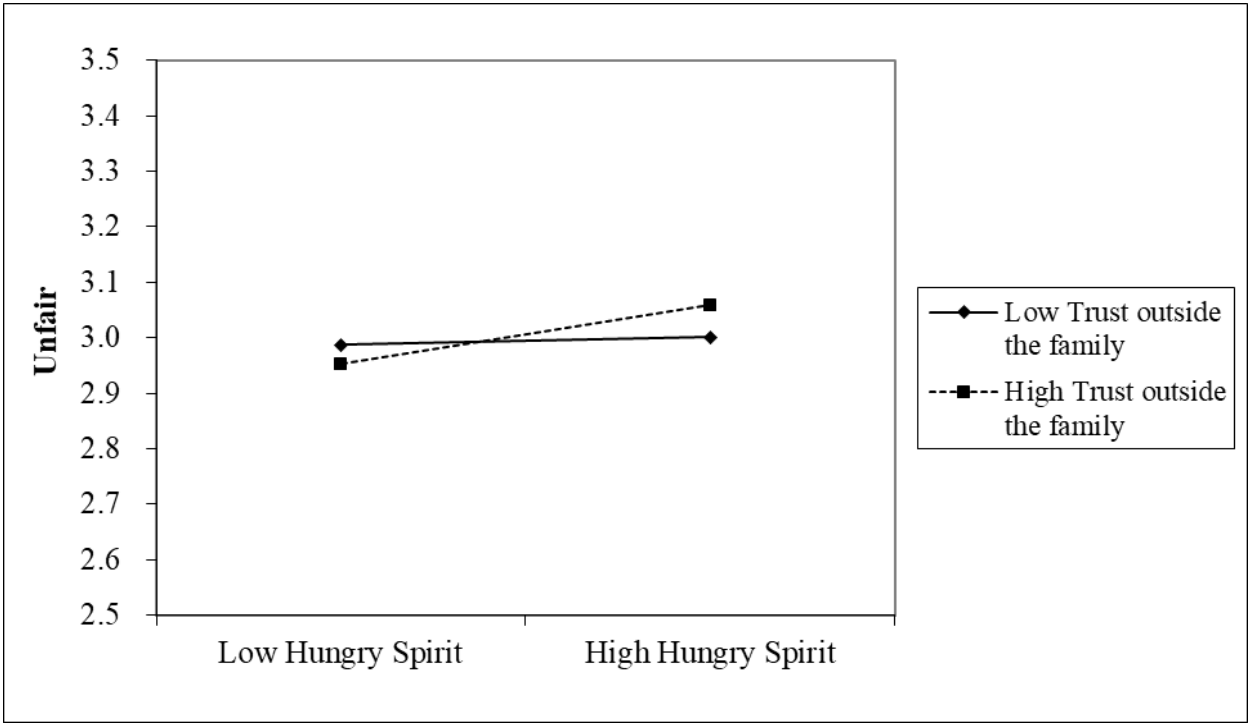

FIGURE 10  The moderating effect of trust outside the family between hungry spirit and unfair

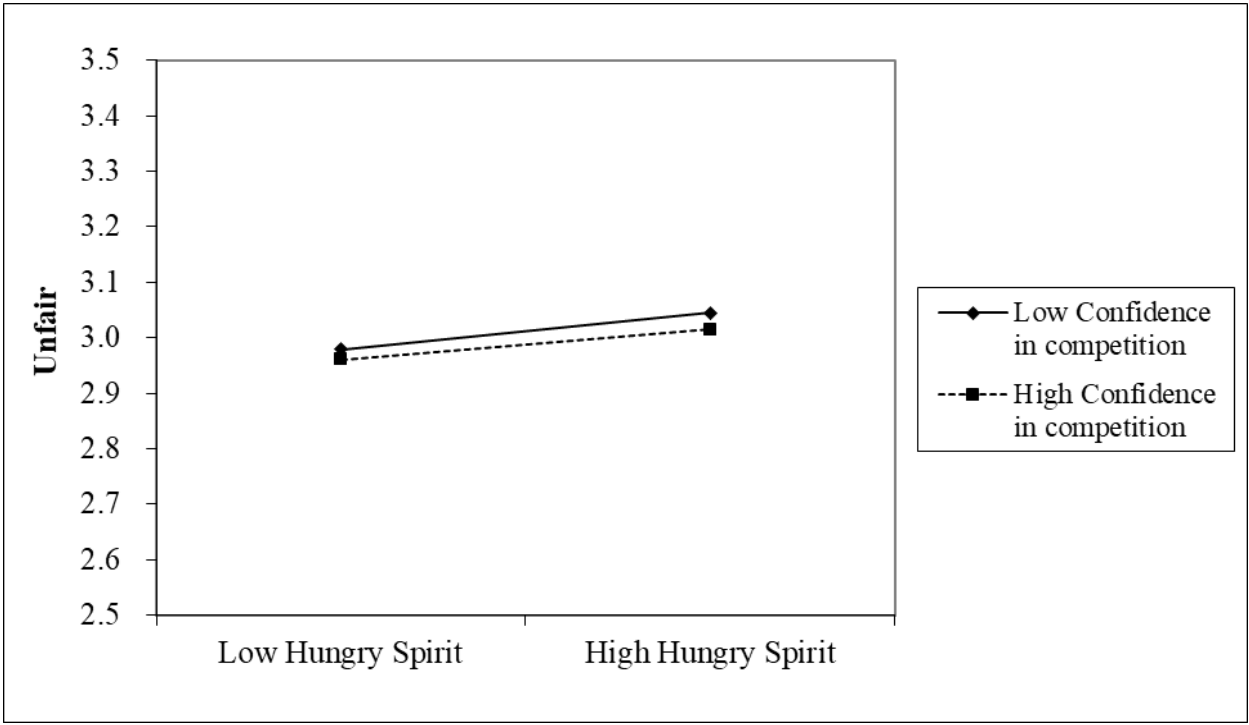

FIGURE 11  The moderating effect of confidence in competition between hungry spirit and unfair

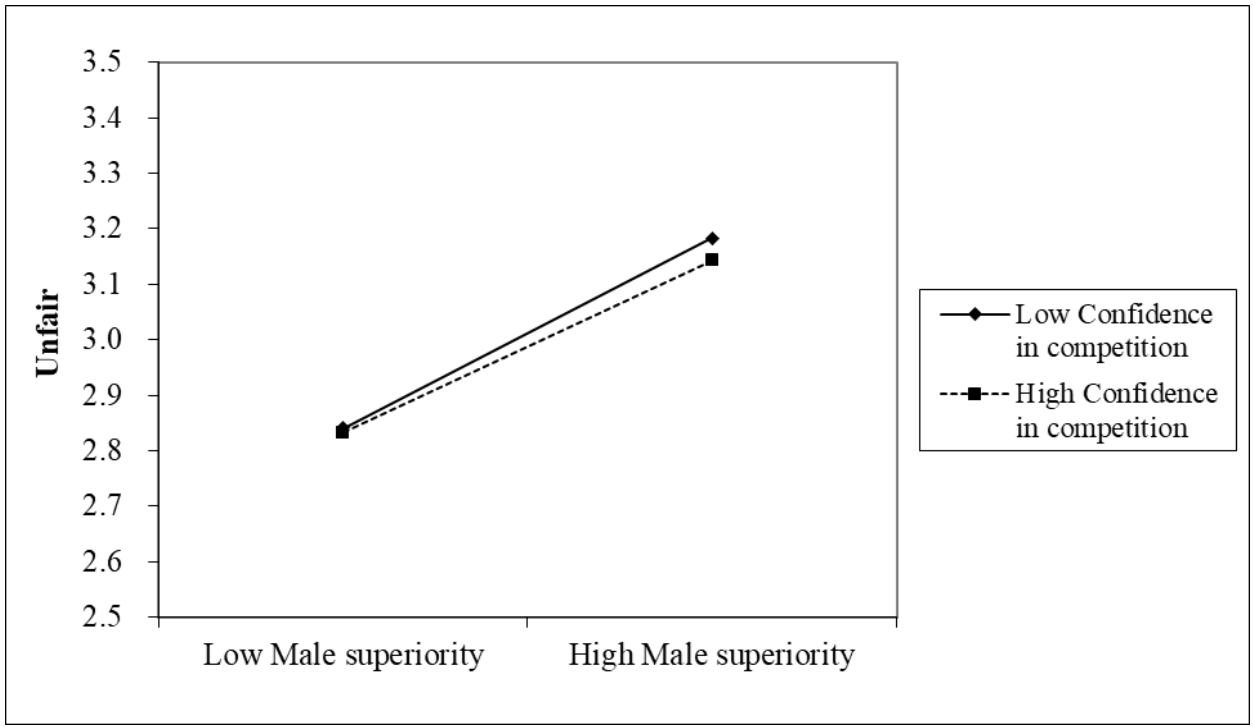

FIGURE 12  The moderating effect of confidence in competition between male chauvinism and unfair

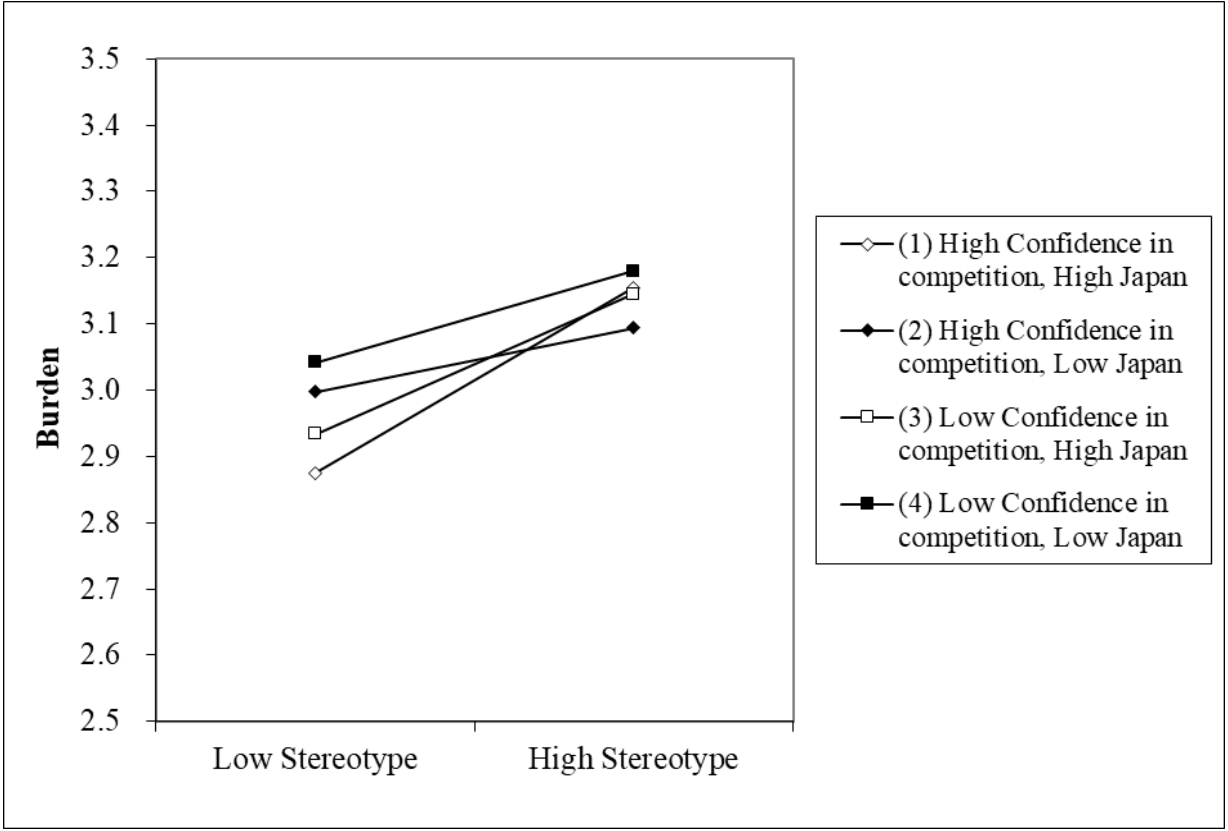

FIGURE 13   The moderating effect of Japan between stereotype and confidence in competition

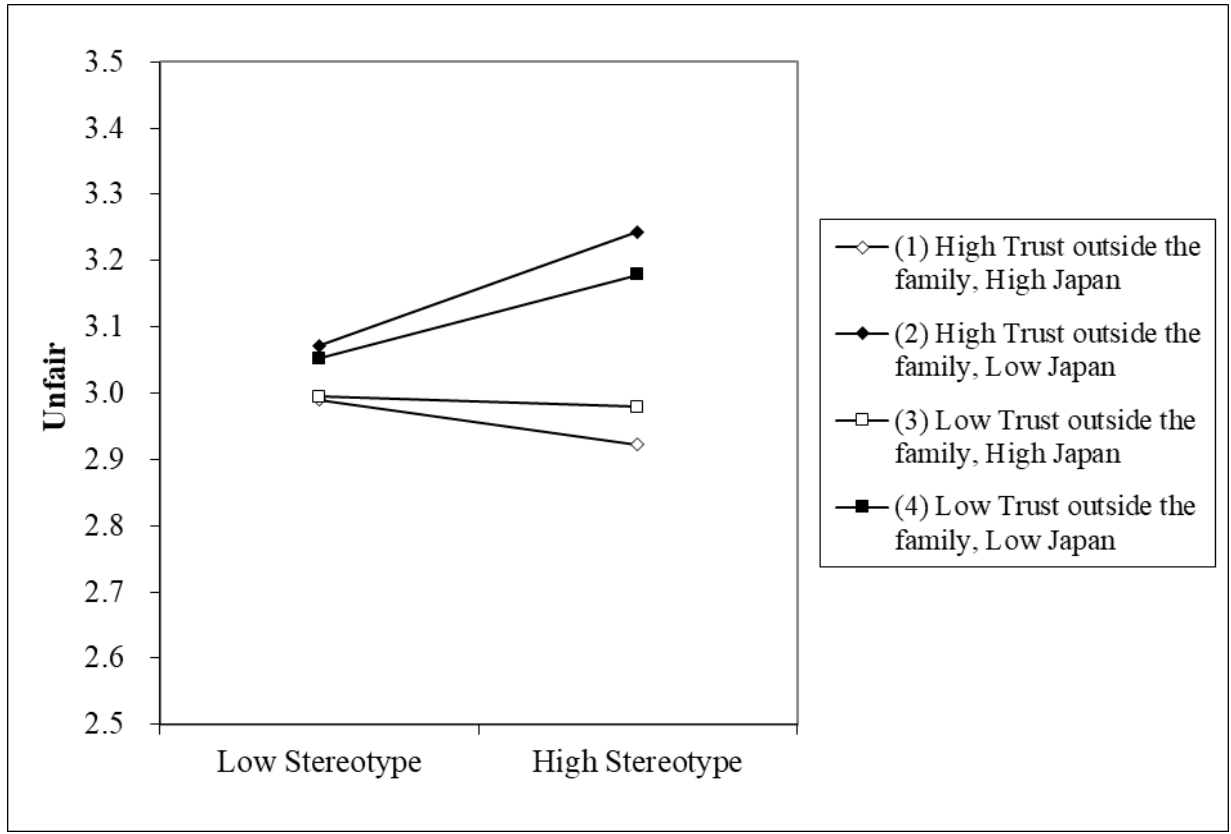

FIGURE 14   The moderating effect of Japan between stereotype and confidence

Discussion

The results of this study confirm the findings of previous studies that stereotypes, hunger, and male chauvinism increase ageism. At the same time, we discovered that altruism, trust within the family, and trust in competition decrease ageism, while trust outside the family increases ageism.

First, consistent with previous studies (Nelson, 2016; Swift et al., 2017), stereotypes were related to ageism. Furthermore, hunger was related to ageism. This is because people with a strong attachment to money are more sensitive to the social burden of older people, and is consistent with previous research that showed that a "market mentality" increases ageism using a dataset that overlaps with this study (Hövermann & Messner, 2023). Furthermore, male chauvinism was related to ageism. This is because people who view a physically masculine, strong, and muscular body as the ideal are more likely to have negative attitudes not only toward women but also toward physically weakened men, which is consistent with previous research (Ng & Lim-Soh, 2021).

However, ageism is not inevitable. Several factors can mitigate the impact of stereotypes, hunger, and male chauvinism on ageism. One is altruism. For example, if we can encourage altruism among individuals who have stereotypes about older people, we can prevent ageism to some extent. The second is trust within the family. As this study showed, even individuals with a high hungry spirit can mitigate the rise of ageism if they have a trusting relationship with their family. However, trusting relationships with friends and acquaintances outside the family strengthen the relationship between stereotypes, hungry spirits, male chauvinism, and ageism. These results are consistent with the arguments of previous research that showed that evaluations of older people are formed completely differently depending on whether they are within the family or outside the family (Hagestad & Uhlenberg, 2005; McPherson et al., 2001; Newman, Faux, & Larimer, 1997; Swift et al., 2017). Thus, a family consisting of age-diverse members is a facilitator for understanding the perspectives of people of different generations, while other general human relationships, which are often composed of people of similar ages, discourage the understanding of people of different generations. The impact of differences in the age composition of groups on outcomes has been overlooked in research on social capital, which assumes that building personal networks contributes to solving all kinds of problems in society (Hagestad & Uhlenberg, 2005). However, in light of research that has pointed out that solidarity among people of similar cultural or racial backgrounds promotes the exclusion of outsiders (Portes, 2009) and research that has shown that ageism is stronger in collectivist cultures than individualist cultures (North & Fiske, 2015), the relationship between age homogeneity and ageism may not have been difficult to predict. In other words, future research may be able to describe or analyze in more detail the path by which trust between age-homogeneous members leads to ageism as a new negative aspect of social capital. This is expected to provide hints for how to implement this technology in society for age-diverse groups, for example.

   Furthermore, the results of this study show that the emergence of ageism can also be prevented by marketization. In other words, as marketization advances and trust in competition increases, the mobility of movement between social classes increases, and ageism declines. In other words, it is not the fate of marketization that leads to ageism, but rather it is suggested that it occurs because the market does not function well and there is no outlet for the hungry spirit, which leads to the older people being scapegoated. In addition, if social mobility is high, the awareness of older people that they have achieved a status commensurate with their abilities will increase among young people, and as a result, the emergence of negative stereotypes such as "older people are those who are reaping the benefits of vested interests" will be prevented. Of course, the existence of such a mechanism cannot be conclusively determined from the analysis of this study, so it should be verified in future research.

   Finally, the results of this study show that there are regional differences in the above trends. In Japan, which had the world's highest aging rate and aging acceleration at the time of the survey, the higher the trust in competition, the more likely people are to feel that older people are a burden due to the influence of stereotypes, a trend different from other countries. This may be because, in Japan, the change in social structure has not kept up with the rapid aging rate, and this may lead to the perception that even if the market is perfect, the gap between generations cannot be bridged (for example, "young people are determined by competition in the free market, but the older people are protected by the legacy of seniority and live comfortably in a different world"). In addition, the more trusting relationships are built outside the family and the more exposed people are to stereotypes, the less they feel that the older people are unfair, which is the exact opposite of the trend in other countries. This may be related to the fact that in Japan, where the aging rate is high, there tends to be a large age diversity in human relationships in general, unlike in other countries. In other words, it suggests that the more intimate relationships people build with groups outside the family, the more likely they are to build trusting relationships with older people, and as a result, the more likely they are to be wary of older people's stereotypes. Therefore, in a highly aged society, we may not need to worry too much about the risk of being reminded of the negative side of social capital, such as increased ageism through involvement with groups outside the family.

Limitation

This study analyzed the factors behind ageism through cross-sectional analysis. Therefore, it is important to note that the results of this study are correlational and do not indicate a causal relationship. It is also important to note that because data from around the world was combined, the unique cultural differences of each country may be underestimated. Future research should verify and develop the results of this study through longitudinal and individual analyses.

Conclusion

As the aging of the world accelerates, it is an urgent task to clarify factors that can prevent ageism. In this study, we used hierarchical multiple regression analysis using data from 40,869 people from 55 countries collected in the World Values Survey Wave 6 to show that, after controlling for demographic factors, stereotypes, hunger, and male chauvinism are related to ageism, and that altruism, trust within and outside the family, and trust in competition moderate the relationship between the independent and dependent variables.

References
Aiken, L.S., West, S.G. , & Reno, R.R. (1991). Multiple Regression: Testing and Interpreting